# BEAM EMITTANCE MEASUREMENTS FOR THE LOW-ENERGY DEMONSTRATION ACCELERATOR RADIO-FREQUENCY QUADRUPOLE[*]

M. E. Schulze, General Atomics, San Diego, CA 92186, USA

J. D. Gilpatrick, W. P. Lysenko, L. J. Rybarcyk, J. D. Schneider, H. V. Smith, Jr., and L. M. Young, Los Alamos National Laboratory, Los Alamos, NM 87545, USA

*Abstract*

The Low-Energy Demonstration Accelerator (LEDA) radio-frequency quadrupole (RFQ) is a 100% duty factor (CW) linac that delivers >100 mA of $H^+$ beam at 6.7 MeV. The 8-m-long, 350-MHz RFQ structure accelerates a dc, 75-keV, 110-mA $H^+$ beam from the LEDA injector with >90% transmission. LEDA [1,2] consists of a 75-keV proton injector, 6.7-MeV, 350-MHz CW RFQ with associated high-power and low-level rf systems, a short high-energy beam transport (HEBT) and high-power (670-kW CW) beam stop. The beam emittance is inferred from wire scanner measurements of the beam profile at a single location in the HEBT. The beam profile is measured as a function of the magnetic field gradient in one of the HEBT quadrupoles. As the gradient is changed the spot size passes through a transverse waist. Measurements are presented for peak currents between 25 and 100 mA.

## 1 INTRODUCTION

The primary objective of LEDA is to verify the design codes, gain fabrication knowledge, understand beam operation, measure output beam characteristics, learn how to minimize the beam-trip frequency, and improve prediction of costs and operational availability for the APT accelerator. The configuration of the LEDA RFQ accelerator is shown in Figure 1. This paper presents the analysis of quad-scan measurements of the output beam from the 6.7 MeV RFQ.

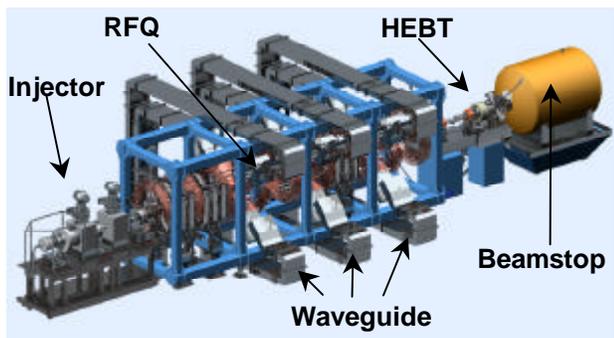

Figure 1. LEDA configuration for RFQ commissioning.

A schematic of the LEDA HEBT [3] showing the location of beamline magnets and diagnostics is given in Figure 2. The function of the LEDA HEBT is to characterize the properties of the beam and transport the beam with low losses to a shielded beamstop. The beamline magnets consist of four quadrupoles and two sets of X-Y steering magnets. The HEBT contains beam diagnostics that allow measurement of pulsed-beam-current, dc-beam-current, and bunched-beam-current as well as transverse centroid, longitudinal centroid (i.e., beam energy from time-of-flight and beam phase), and transverse beam profile (wire scanner and video fluorescence) [4].

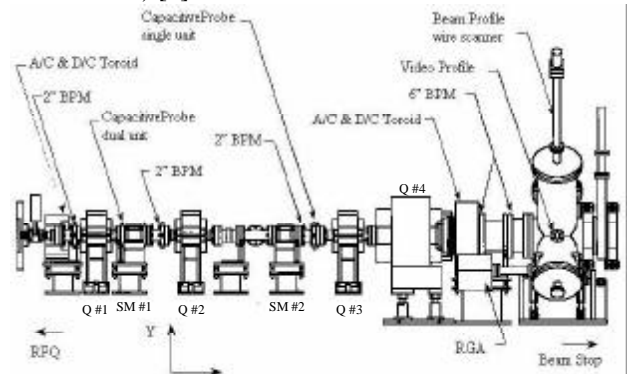

Figure 2. Layout of HEBT beamline optics and diagnostics. Beam direction is from left to right.

Quad-scan measurements were made using only the first two quadrupole magnets. For the horizontal scan, the first quadrupole in the HEBT was fixed while the second quadrupole was varied over a range between 4.7 T/m and 10.7 T/m. For the vertical scan, the second quadrupole in the HEBT was fixed while the first quadrupole was varied over a range between -7.0 T/m and -13.0 T/m. In both cases the two downstream quadrupoles were off. The resulting beam size measured at the wire profile monitor passes through a minimum as the gradient is varied. In each scan, beam profile measurements were typically made at nine different settings. Profile measurements were performed at beam peak currents of 25, 50, 75 and 94 mA. Many of these measurements were repeated over a period of one month.

[*] Work supported under DOE Contract DE-AC04-96AL99607

## 2 ANALYSIS

The analysis of the quad scan data utilized the beam optics code LINAC [5] and only addresses the rms properties of the distribution. The Twiss parameters $\alpha$, $\beta$, and $\epsilon$ at the exit of the RFQ were adjusted to fit the rms widths of the beam distributions taken during the quad scans. A six dimensional waterbag distribution was assumed although the analysis is generally independent of the distribution. Initially, the Twiss parameters were adjusted to fit the predicted distribution from RFQ simulations (PARMTEQM or nominal) for a 94 mA beam current. Beginning with these Twiss parameters, the sensitivity to $\epsilon$, $\alpha$, and $\beta$ was analysed at 94 mA for the two transverse planes. No attempt was made to study the longitudinal properties of the distribution.

The data sets (rms width vs. quadrupole gradient) were combined and analyzed to obtain a polynomial fit as shown in Figures 3a and 3b. The data sets represent measurements made on three occasions.

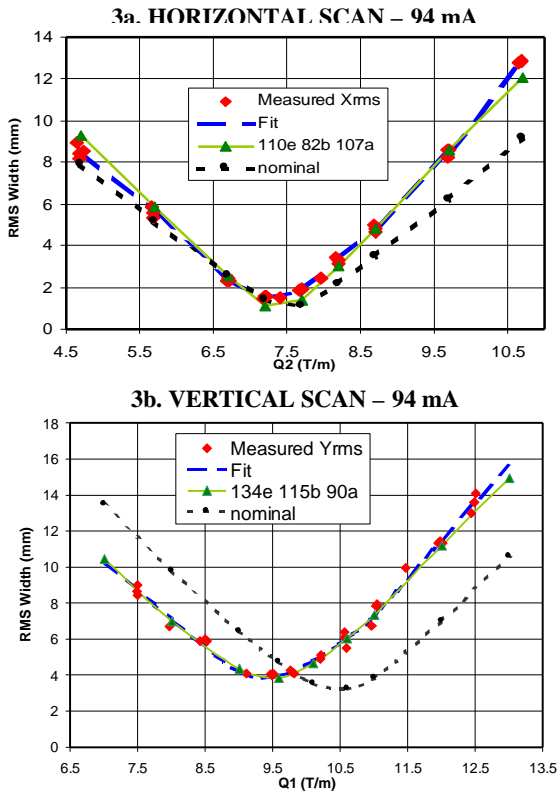

Figures 3a and 3b. Measured values of the rms beam width (red triangles) and corresponding fits (long blue dashes) at 94 mA. The black curves (small dashes) show the nominal RFQ output beam and the green (solid) curves shows the best fit to the data using LINAC.

From the polynomial fits to the data the rms beam width was calculated at nine gradient settings. The LINAC code was then run at each of these gradient settings for a specific set of Twiss parameters. The Twiss parameters were then optimized by minimizing the sum of the squared errors (SSE) between the fit and the results of the nine runs. Typically, the SSE was less than 1.0 mm$^2$. Figures 3a and 3b show the fits to the data in the horizontal and vertical planes at 94 mA. The black curve labelled nominal represents the RFQ beam simulation from PARMTEQM [6]. The green curve represents the best fit by adjusting the Twiss parameters according to the label (e.g. 110e 82b 107a means 1.10 times the nominal emittance, 0.82 times the nominal beta and 1.07 times the nominal alpha). These parameters indicate that the horizontal beam size is about 5% larger at the RFQ exit with a 20% higher divergence than that predicted by PARMTEQM. In comparison, the vertical beam size is about 25% larger at the RFQ exit with a divergence consistent with that predicted by PARMTEQM.

The same analysis was performed on data taken at 25 mA. The results are shown in Figures 4a and 4b. The analysis shows the horizontal beam emittance to be about 90% of the 94 mA emittance predicted by PARMTEQM with the horizontal beam size about 12% smaller at the RFQ exit and the divergence about 4% higher. The vertical beam distribution has an emittance close to the design prediction at 94 mA while the beam size and divergence are about 75% of the nominal design.

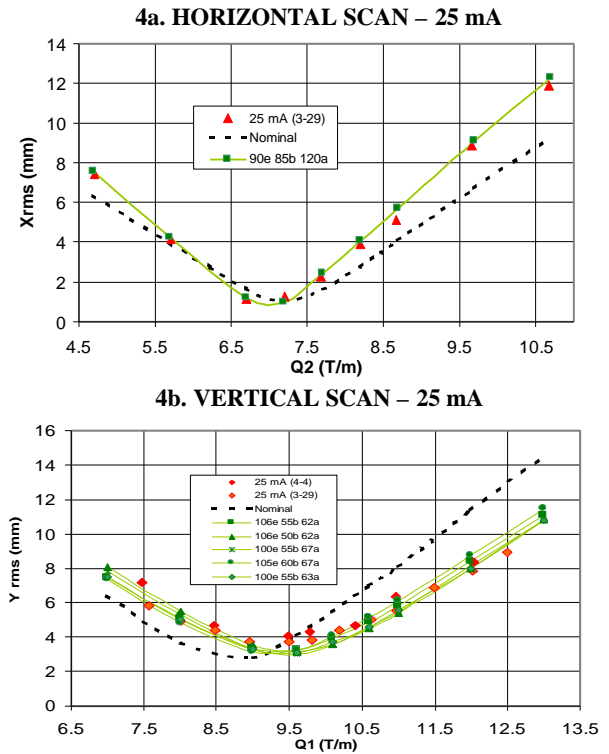

Figures 4a and 4b. Measured values of the rms beam width (red triangles) and corresponding fits (long blue dashes) at 25 mA. The black curves (small dashes) show the nominal RFQ output beam and the green (solid) curves shows the best fit to the data using LINAC

It should be noted that the beam measurements made at 25 mA were not made with the injector matched to the RFQ. Figure 4b shows many different fits to the vertical rms spot size that produces similar SSE's. The overall variation in the Twiss parameters is only about 5%.

A similar analysis has been performed for a quad scan taken at 75 mA peak current. Table 1 summarizes these results as well as the results at 25 and 94 mA.

Table 1: Normalized Beam Emittance and Twiss Parameters from Quad Scan Analysis

| Current | Nominal | 94 mA | 75 mA | 25 mA |
|---|---|---|---|---|
| $\varepsilon_{xn}$ $\pi$(mm-mrad) | 0.229 | 0.253 | 0.216 | 0.207 |
| $\alpha_x$ | 1.671 | 1.785 | 1.913 | 1.919 |
| $\beta_x$ meters | 0.436 | 0.357 | 0.379 | 0.371 |
| $\varepsilon_{yn}$ $\pi$(mm-mrad) | 0.234 | 0.314 | 0.268 | 0.258 |
| $\alpha_y$ | -2.750 | -2.483 | -2.479 | -1.711 |
| $\beta_y$ meters | 0.778 | 0.892 | 0.649 | 0.427 |

## 3 DISCUSSION

During commissioning it was observed that the RFQ fields had to be increased 5-10% above the design value for optimal transmission [7]. The effect of this higher field has been studied and does not result in significant changes in the predicted PARMTEQM beam distributions.

The effects of beam neutralization were investigated for the 94 mA data. A small reduction in the SSE was observed in the horizontal quad scan data consistent with about 10% neutralization. Analysis of the vertical quad scan data with 10% neutralization resulted in Twiss parameters that were within 5% of those presented in Table 1.

A rigorous error analysis was not performed in the analysis of the quad scan data. The different errors that contribute to the analysis uncertainty have been estimated. These errors result from beam jitter (1.0 mm typically), measurement reproducibility (0.5 mm), background subtraction (< 0.5 mm), analysis uncertainty (5%), and quadrupole gradient fluctuations (1-2%). The effect of beam jitter is most pronounced in the minimum of the horizontal quad scan data where it is observed that the simulations are consistently lower than the measurements.

A systematic uncertainty in the analysis results from the use of the LINAC code to infer the Twiss parameters. LINAC includes non-linear space-charge effects, which are essential to modelling the beam transport. TRACE-3D [8], which includes only linear space charge, was found to be inadequate in describing the beam distributions. Beam profiles produced by LINAC were generally in good agreement with the measured beam profiles for the lower quadrupole gradients before the minimum. After the minimum the measured beam distributions exhibited significant tails which are not well reproduced by LINAC even though the rms widths are in close agreement. One difficulty in reproducing the exact shape of the distribution is attributed to the large aspect ratio (>5) between the horizontal and vertical beam widths when the beam is at a waist [9,10]. The details of the beam distribution from the RFQ are also unknown.

IMPACT, a more sophisticated PIC beam-optics code, has been used to analyze the quad scan distributions [9]. This analysis gives essentially the same rms widths as LINAC but with significantly better agreement in predicting the shape of the profiles.

## 4 SUMMARY

The rms output beam parameters from the 6.7 MeV LEDA RFQ have been inferred from quad scan measurements using the LINAC beam optics code. Analysis of the data presented in this paper continues. We are now preparing to intentionally introduce and measure the beam halo in a 52-magnet FODO lattice [11]. This measurement will also allow for an independent measure of the beam emittance.